# PRACTICAL LARGE-SCALE SPATIO-TEMPORAL MODELING OF PARTICULATE MATTER CONCENTRATIONS[1]

By Christopher J. Paciorek, Jeff D. Yanosky, Robin C. Puett, Francine Laden and Helen H. Suh

*Harvard University, Harvard University, University of South Carolina, Harvard University and Brigham and Women's Hospital and Harvard University*

The last two decades have seen intense scientific and regulatory interest in the health effects of particulate matter (PM). Influential epidemiological studies that characterize chronic exposure of individuals rely on monitoring data that are sparse in space and time, so they often assign the same exposure to participants in large geographic areas and across time. We estimate monthly PM during 1988–2002 in a large spatial domain for use in studying health effects in the Nurses' Health Study. We develop a conceptually simple spatio-temporal model that uses a rich set of covariates. The model is used to estimate concentrations of $PM_{10}$ for the full time period and $PM_{2.5}$ for a subset of the period. For the earlier part of the period, 1988–1998, few $PM_{2.5}$ monitors were operating, so we develop a simple extension to the model that represents $PM_{2.5}$ conditionally on $PM_{10}$ model predictions. In the epidemiological analysis, model predictions of $PM_{10}$ are more strongly associated with health effects than when using simpler approaches to estimate exposure.

Our modeling approach supports the application in estimating both fine-scale and large-scale spatial heterogeneity and capturing space–time interaction through the use of monthly-varying spatial surfaces. At the same time, the model is computationally feasible, implementable with standard software, and readily understandable to the scientific audience. Despite simplifying assumptions, the model has good predictive performance and uncertainty characterization.

**1. Introduction.** A growing body of evidence documents chronic health effects of air pollution. Two prospective studies of mortality have been par-

Received January 2008; revised August 2008.
[1]Supported by EPA STAR R83-0545-010 and the Harvard/EPA PM Center, EPA STAR R-832416-010.

ticularly influential in demonstrating these effects, significantly affecting environmental policy: the Harvard Six Cities Study [Dockery et al. (1993)] and the American Cancer Society (ACS) Study [Pope et al. (1995, 2002)]. These studies showed heightened individual risk of mortality in more polluted metropolitan areas based on associations with time-invariant estimates of fine particulate matter, $PM_{2.5}$ (PM less than 2.5 $\mu$m in aerodynamic diameter), which varied at the city level. The ACS study averaged over all monitors in each metropolitan area to estimate exposure, while the Six Cities study focused on smaller cities and used a central site monitor, recruiting participants living near the site. More recently, spatial models that account for fine-scale heterogeneity have been used to estimate variability in exposure within metropolitan areas to link to health effects [Jerrett et al. (2005), Künzili et al. (2005)].

Atmospheric particulate matter (PM) originates from a variety of stationary and mobile sources and may be directly emitted (primary emissions) or formed in the atmosphere by transformation of gaseous emissions (secondary emissions). Most monitoring in the US concerns $PM_{2.5}$ and $PM_{10}$ (PM less than 10 $\mu$m in aerodynamic diameter), with coarse PM, $PM_{10-2.5}$, generally measured by difference. While combustion sources, which account for the bulk of anthropogenic PM emissions to the atmosphere, typically lead to the formation of $PM_{2.5}$ through primary and secondary emissions, mechanical grinding and crushing activities typically lead to primary emissions of coarse mode particles. The kinetics of the atmospheric transformation of precursor gases play a key role in determining the spatial distribution of the components of fine mode PM, with more reactive species exhibiting greater spatial heterogeneity than more stable species. For example, sulfur dioxide emissions are oxidized slowly relative to many other compounds and, as a result, secondary ammonium sulfate particles are spatially homogenous across large distances. In contrast, coarse mode particles are removed from the atmosphere more quickly, by gravitational settling and other processes, and are therefore typically more spatially heterogenous than fine mode particles [Burton et al. (1996)].

Our work is part of a larger project analyzing the associations between health outcomes and PM exposure in a large cohort study, the Nurses' Health Study (NHS) [Colditz and Hankinson (2005)], in which we aim to use more precise exposure estimates than have been used in previous studies to increase power and reduce potential bias from measurement error. The NHS was established in 1976 as a cohort of 121,700 female registered nurses between the ages of 30 and 55, living initially in 11 large states, primarily in the northeast US. For the health analyses, interest lies in estimating concentrations every month for 1988–2002 for the northeast US, with extension to much of the rest of the country ongoing. The size of the domain is particularly salient; our goal is to build a model that captures spatio-temporal patterns and can be feasibly fit over the entire northeast US domain, making monthly predictions at approximately 70,000 locations over 15 years based on monitoring data at over 900 locations. Two particular aspects of the space–time structure motivate our model choice. First, within the context of the large spatial domain, we seek to estimate fine-scale spatial heterogeneity to more precisely estimate potential exposure to PM than from central site proxies. Second, we seek to estimate space–time interaction: monthly heterogeneity is of interest because the health modeling attempts to understand the relevant time window of exposure to PM associated with health outcomes, using moving averages to estimate exposure in the previous $X$ months, where $X$ varies between analyses. Purely temporal heterogeneity is of less interest (and is, in any event, well-characterized with hundreds of monitors at each point in time) because such variability is likely to be confounded with other time-varying factors that affect health outcomes and is generally conditioned out of the epidemiological model. Puett et al. (2008) present initial health analyses using the exposure estimates from our modeling.

Statistical space–time modeling holds promise for better resolving spatial and temporal heterogeneity in concentrations, both to estimate exposure for health studies and to characterize patterns of PM for understanding variability and attainment of environmental standards. The inclusion of covariates in the mean structure may help to resolve spatial heterogeneity at a resolution not possible from purely spatial or spatio-temporal smoothing of data. We make use of the recent statistical and computational innovation in additive modeling with smooth terms [Kammann and Wand (2003), Ruppert et al. (2003), Wood (2006)]. There has been much recent statistical work to model spatio-temporal heterogeneity in PM [e.g., Kibria et al. (2002), Daniels et al. (2006)]. Calder (2008) jointly models daily $PM_{10}$ and $PM_{2.5}$ for one year in Ohio using a process convolution approach for the spatial structure and a dynamic linear modeling approach to capture temporal evolution. Sahu et al. (2006) model weekly PM in the midwest US using land use covariates in the mean with separate separable space–time covariances for a rural, background process and an urban process. Smith et al. (2003) took an approach most similar to that used here, modeling weekly PM in the southeast US based on a temporal term, a spatial term modeled using a thin plate spline and a land use covariate, and spatio-temporal residuals modeled independently by kriging each week separately. Such methods have not yet made their way into the environmental science literature, but so-called land use regression is very popular, generally taking the form of linear regression of a pollutant on spatially-varying covariates reflecting land use and emissions [Briggs et al. (2000), Brauer et al. (2003)].

While the statistical efforts described above demonstrate the wide array of potential model structures, most of the work focuses on domains of limited size, whereas chronic epidemiological studies often require estimation in large domains for adequate statistical power to estimate small risks. Complicated models that more fully account for the rich space–time structure of the data can be difficult to specify and fit. When the goal is prediction, one might consider simpler specifications that adequately predict the exposure of interest, potentially with little loss of predictive power so long as key aspects of the data structure are represented. Here we present a computationally-tractable space–time model for PM over a large space–time domain, serving as the core exposure estimation used in an extensive epidemiological analysis. Companion papers [Yanosky et al. (2008a, 2008b)] focus on the scientific results for $PM_{10}$ and $PM_{2.5}$ respectively; here we focus on the statistical issues. The model accounts for important factors that improve prediction and uncertainty characterization of PM, while retaining computational feasibility, interpretability, and reliance on standard software for applied use.

We describe the PM data and covariates in Section 2. In Section 3 we describe our two-stage spatio-temporal model for dense data fit via backfitting with gam in R. We then report the predictive performance of the model for $PM_{10}$ for 1988–2002 and $PM_{2.5}$ and coarse PM for 1999–2002 and discuss residual and prediction error diagnostics. Finally, we use the predictions and simpler alternatives in an epidemiological analysis. In Section 4 we develop a simple conditional model, built on the core model, to predict $PM_{2.5}$ from $PM_{10}$ when $PM_{2.5}$ data are sparse, which we apply to $PM_{2.5}$ for the years 1988–1998. For this, we discuss techniques for using airport visibility information as a proxy for $PM_{2.5}$, using a stochastic EM algorithm to deal with interval censoring.

**2. Data.** We provide a synopsis of the data used; see Yanosky et al. (2008a, 2008b) for a more extended description. The spatial domain is the northeastern United States, but we include monitors from neighboring states to avoid boundary effects (Figure 1). We model monthly concentrations for the years 1988–2002. The data are available in the supplemental material [Paciorek et al. (2009)].

2.1. *Monitoring data.* $PM_{10}$ and $PM_{2.5}$ mass concentrations, measured using US Environmental Protection Agency (EPA)-approved Federal Reference Method or equivalent methods, were obtained from the EPA Air Quality System (AQS) database (Figure 1). We also obtained $PM_{10}$ and $PM_{2.5}$ mass concentrations from the IMPROVE (Interagency Monitoring of PROtected Visual Environments) network, whose sites are located in national parks and wilderness areas, and additional data from the Stacked Filter Unit network (a predecessor to IMPROVE), the Clean Air Status and Trends (CASTNet) network, and Harvard research studies. There are 922 $PM_{10}$ and 498 $PM_{2.5}$ sites in the domain, with the sites providing either 24-hour average or hourly average PM concentrations, with most $PM_{10}$ ($PM_{2.5}$) monitors operating one in six (three) days, but some operating every day. No sites report for the entire time period (the range of number of observations per month is 135–475 for $PM_{10}$ and 8–446 for $PM_{2.5}$ with 8–25 before 1999 and 107–446 after 1998). Missing daily observations within a period in which the site is operating generally occur for reasons unrelated to the levels of pollution, mainly equipment problems, maintenance, and addition or removal of a site, and there is no evidence of any patterns of missingness, so the missing values can be considered missing completely at random (MCAR). From the available daily averages, we calculated monthly averages, excluding site-month pairs with fewer than four daily values or with more than one-third of scheduled observations missing, to avoid data that may be unrepresentative of the pollution at the site in a given month. As a sample of the spatial domain, we assume that pollution levels at locations without monitors are missing at random (MAR) and therefore ignorable, conditional on the covariates we include in the model to represent local conditions (see Section 5 for additional consideration of this assumption).

2.2. *GIS and meteorological covariate data.* We used a Geographical Information System (GIS) to generate many potential predictors, based on government and other databases. The non-time-varying covariates were as follows: distances to the nearest road within four road size classes; particulate point source emissions within 1 and 10 km buffers; the proportion of urban land use within 1 km; elevation; and block group, tract, and county population density from the 1990 US Census.

Meteorological variables at WBAN (Weather Bureau Army Navy) and other weather stations were obtained from the National Climatic Data Center. The variables considered were temperature, precipitation, barometric pressure, and wind speed, with hourly values averaged to the month at each station. We spatially smoothed the meteorological variables using a simple GAM with a two-dimensional penalized spline smooth term for the geographic locations to provide estimated values at all locations in the study region for each month. County area source emissions for each year were also considered as time-varying covariates.

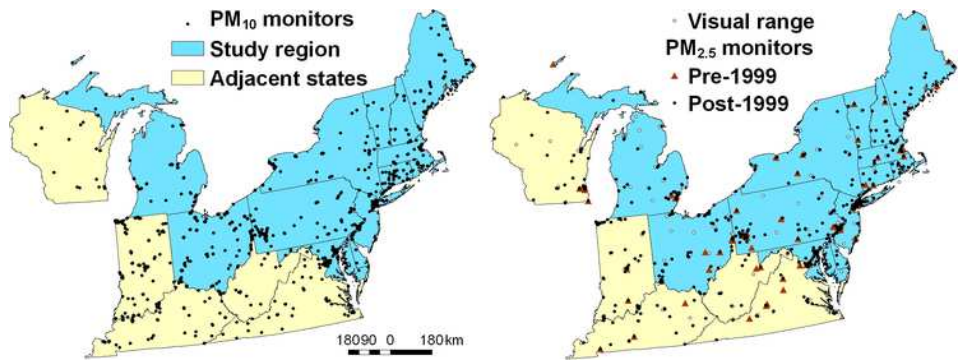

FIG. 1. *Map of $PM_{10}$ (left) and $PM_{2.5}$ (right) monitoring locations in the study region and adjacent states. For $PM_{2.5}$, monitors are grouped by availability before 1999.*

More details on the covariates and the covariate selection process are available in Yanosky et al. (2008a, 2008b). Here we take the set of covariates to be fixed and focus on the core model and results, as well as alternative model specifications.

### 3. An additive space–time model for PM using dense monitors.

3.1. *Core spatio-temporal model.* We first present the core model, which is applied separately to $PM_{10}$ for 1988–2002 and $PM_{2.5}$ for 1999–2002. We chose to fit separate models for $PM_{10}$ and $PM_{2.5}$ rather than model jointly because $PM_{2.5}$ and $PM_{10}$ monitors are relatively dense, often (about half the monitors) co-located at the same site, and have similar siting patterns, with more monitors in more densely-populated areas. As a result, we assume that measurements of one provide limited information about the other, once we condition on the data available for the pollutant of interest and on the covariates. We feel that the computational expense and increase in model complexity of a bivariate outcome model do not justify a joint model. However, there are many locations with only one type of monitor and no nearby monitors. For 1999–2002 this occurs primarily as sites with only a $PM_{2.5}$ monitor (except for Ohio and northern Maine) [Paciorek et al. (2009), Figure S1], so a joint model might provide some improvement in predictive performance for $PM_{10}$.

3.1.1. *Model structure.* We propose the following two-stage model. The first stage models log PM at site $i$ and month $t$,

$$(3.1) \qquad \log PM_{it} = y_{it} \sim \mathcal{N}\left(\mu_i + \sum_k h_k(x_{kit}) + g_t(s(i)), \sigma_t^2\right),$$

using fixed effects for each site, $\mu_i$, to account for space-only variation while accounting for space–time interaction based on smooth regression functions, $h_k(\cdot)$, of time-varying covariates and a time-varying residual spatial surface, $g_t(\cdot)$, for each month, $t \in \{1, \ldots, T\}$, independent between months, where $s(i)$ is the spatial location of the $i$th site. $\sigma_t^2$ is the homoscedastic (in space), monthly-varying residual variance. A more sophisticated model that is part of our ongoing work decomposes this residual variance into several variance components, representing fine-scale heterogeneity, instrument error, and (heteroscedastic) variability from not having everyday measurements to estimate monthly average PM.

The second stage models the predictable component of the fitted site-specific terms, $\hat{\mu}_i$, using smooth regression functions of time-invariant covariates, $f_j(\cdot)$, and a spatial surface, $g_\mu(\cdot)$,

$$\hat{\mu}_i = g_\mu(s(i)) + \sum_j f_j(z_{ji}) + b_i, \tag{3.2}$$

where the residuals, $b_i$, which are taken to be normal with variance $\sigma_\mu^2$, represent unexplained site-specific (fine-scale) variability. An alternative to our two-stage model would be a true generalized additive mixed model (GAMM) resulting from substituting (3.2) into (3.1), but the time-varying spatial surfaces make this difficult to fit.

The model allows us to estimate spatial patterns for individual months, while borrowing strength across months to estimate time-invariant regression effects. The smooth regression terms flexibly account for possible nonlinearities; interpretation of individual terms needs to recognize the possibility of concurvity. More details on covariate selection are given in Yanosky et al. (2008a, 2008b). By fitting an effect for each site, $\mu_i$, we account for correlation across time at each site, thereby limiting overfitting of the time-invariant covariates and spatial effects. However, our model takes each replicate at a site as a separate observation for the time-varying terms, $h_k(\cdot)$, rather than acknowledging the repeated sampling. To avoid estimating these terms to be less smooth than scientifically plausible, we enforced additional smoothness using the 'sp' argument in `gam`.

We specify stationary spatial structures for $g_\mu(\cdot)$ and $g_t(\cdot)$ through the use of penalized thin plate splines. Note that the presence of the covariates, in particular, a variety of land use covariates that distinguish urban and rural areas, helps justify the stationarity assumption, but we explore the presence of additional nonstationarity in the supplementary material [Paciorek et al. (2009), Section S1]. Our assumption seems particularly defensible for $PM_{2.5}$, whose sources are more regional in nature.

One important simplification in the model that eases fitting is the assumption of independence between the residual spatial surfaces, $g_t(\cdot)$, for each month. In Section 3.2.3 we show that while there is some residual temporal autocorrelation, there is no apparent residual spatial autocorrelation. This indicates that the systematic variation not captured in the model is variation in time and not in space. Since our predictive modeling is for unobserved locations in space and not new points in time, modeling the temporal structure would be very unlikely to improve our predictions. We justify this theoretically as follows. Assume the following conditions: (1) a space–time covariance model with separable structure, (2) normality, (3) the same locations sampled at every time point, and (4) all error (the nugget) assumed to be local heterogeneity rather than instrument error. The space–time kriging prediction under these assumptions, assuming mean zero for simplicity, is $(C_t \otimes C_{21})(C_t \otimes C_{11})^{-1}Y = (C_t C_t^{-1}) \otimes (C_{21} C_{11}^{-1})Y = I_T \otimes (C_{21} C_{11}^{-1})Y$. This does not involve the temporal covariance, $C_t$, but only the spatial covariance between prediction and observed sites, $C_{21}$, and the spatial covariance among observed sites, $C_{11}$. Thus, the best prediction for a new set of locations at a given time is based only on measurements from that same time, with the predictions conditionally independent of data at other times given the data at the time of interest. Similar calculations demonstrate that the kriging variances (but not the covariances) are based only on the spatial covariance structure. Although our model (3.1)–(3.2) is not a kriging model, the monthly averages still contain some instrument error, and there are some missing observations, this reasoning suggests that we might exclude temporal correlation from the model without drastic consequences. The true covariance is presumably not separable, but at the scale of the month, for which atmospheric transport and dynamics can reasonably be ignored, this assumption seems more reasonable than it would be for data at a finer time scale, such as daily data.

3.1.2. *Model fitting.* The gam function in the mgcv package in R is a convenient tool for additive modeling that performs multiple penalty optimization for multiple smooth terms without the need for backfitting [Wood (2003, 2004, 2006)]. However, in (3.1)–(3.2), the presence of spatial terms specific to monthly subsets of the data, $g_t(\cdot)$, prevents us from fitting the model all at once. We fit the first stage by calling gam in a backfitting procedure. We start by calling gam to fit the regression portion of the model with fixed effects for site and regression smooths of the time-varying effects. We then iterate between the following steps: (1) separately fit the spatial terms for each month to the regression model residuals, $y_{i,t} - \hat{\mu}_i - \sum_k \hat{h}_k(x_{kit})$, and (2) fit the regression portion of the model to the spatial residuals, $y_{it} - \hat{g}_t(s(i))$, iterating until convergence. We then fit the second stage model using a single call to gam with the estimated site-specific fixed effects, $\hat{\mu}_i$, as the outcome.

The basic code used in fitting the model is available in the online supplementary material [Paciorek et al. (2009)].

With approximately equal sample sizes at each site, which is generally true for our data, assuming homoscedasticity in the second stage model seems reasonable. However, with varying sample sizes (and we did have a few sites with few samples), one would want to use a heteroscedastic second-stage model with variance, $\sigma_\mu^2(1 + \kappa \widehat{\text{Var}}(\hat{\mu}_i))$, where the first component accounts for the variance present from unexplainable site-specific heterogeneity and the second for uncertainty in the site-specific estimates, $\hat{\mu}_i$, from finite sampling. As a sensitivity analysis, we used the gamm function [Wood (2006)], which combines the functionality of gam and lme, to fit a weighted second-stage model with variances proportional to $(1 + \kappa \widehat{\text{Var}}(\hat{\mu}_i))$, defining our own variance function and estimating $\sigma_\mu^2$ and $\kappa$. For $PM_{10}$, predictive accuracy decreased slightly, while for $PM_{2.5}$ it increased slightly compared to the homoscedastic model. In both cases the spatial term, $g_\mu(\cdot)$, was estimated to be less wiggly under the heteroscedastic model that discounts those $\hat{\mu}_i$ that are less certain. Coverage of prediction intervals, for reasons that are unclear, was lower under the heteroscedastic model.

3.1.3. *Prediction and uncertainty estimation.* Prediction is straightforward, requiring only the covariate values at the prediction locations and times of interest. To predict on the original scale, one can simply exponentiate the predicted value, which while not unbiased, is 'median-unbiased,' or one could use a bias-corrected prediction, $\widehat{PM}_{it} = \exp(\hat{Y}_{it} + \frac{1}{2}\widehat{\text{Var}}(\hat{Y}_{it}))$ [Schabenberger and Gotway (2005), pages 268–269].

We estimate prediction uncertainty by summing across the uncertainty of the components in (3.1)–(3.2),

$$\widehat{\text{Var}}(\hat{Y}_{it}) \approx \widehat{\text{Var}}\left[\hat{g}_\mu(s(i)) + \sum_j \hat{f}_j(z_{ji})\right] + \hat{\sigma}_\mu^2$$
$$(3.3) \qquad\qquad + \sum_k \widehat{\text{Var}}(\hat{h}_k(x_{kit})) + \widehat{\text{Var}}[\hat{g}_t(s(i))] + \hat{\sigma}_t^2.$$

While gam accounts for dependence in the uncertainty of multiple additive function components within any given model fit, indicated by the first term in (3.3), our uncertainty estimate does assume independence between the time-invariant and time-varying parts of the model and between the monthly spatial surfaces and the time-varying covariate smooth terms. This was necessary because these standard errors were obtained from separate model fits in the backfitting approach. Also, because the first stage model is fit with monitor-specific intercepts, to estimate the uncertainty associated with the time-varying smooth terms, $\hat{h}_k(x_{kit})$, we needed to extract the individual variances for each of these smooth terms. Ignoring the dependence between the various components of uncertainty is a simplification, but given that our coverage results show only slight undercoverage and that the majority of the uncertainty is contributed by $\hat{\sigma}_\mu^2$ and $\hat{\sigma}_t^2$, we feel this is reasonable. To roughly estimate uncertainty in long-term averages, we assume independent contributions from the time-varying components at different times, which is likely to underestimate uncertainty [Stein and Fang (1997)]. This is partially mitigated by inclusion of the time-invariant terms that account for common components of uncertainty that do not decrease with temporal averaging.

Our inclusion of the variance components, $\sigma_\mu^2$ and $\sigma_t^2$, makes the assumption that the residual variability is fine-scale heterogeneity and not instrument error [as defined in Cressie (1993), page 59], because monitoring is quite accurate. These variance components account for prediction uncertainty at new locations on top of the contributions from functional uncertainty. The assumption of limited instrument error seems particularly justifiable for $\sigma_\mu^2$, which is based on repeated measurements at locations. In ongoing work we are exploring models that more carefully decompose the variance, making use of co-located monitors to estimate an instrument error variance component.

To estimate standard errors on the original scale, one can use the delta method, $\widehat{\mathrm{Var}}(\exp(\hat{Y}_{it})) \approx \widehat{\mathrm{Var}}(\hat{Y}_{it})\exp(\hat{Y}_{it})^2$, but use of confidence intervals would avoid the need for this approximation.

3.1.4. *Assessing generalizability using cross-validation.* To assess the accuracy of the point predictions and our quantification of uncertainty, we take a three-way cross-validation approach [Draper and Krnjacic (2006)] that accounts for our extensive model selection efforts [see Yanosky et al. (2008a, 2008b)]. We divide the monitors in the core states (excluding the boundary states) at random into ten groups of approximately equal size, following Hastie et al. (2001). One group is held out as a test set and is not used at all for model selection to assess whether the selection process itself has resulted in overfitting. The remaining nine groups are used in nine-fold cross-validation to compare models. For each set, a given model is fit using the remaining eight (training) sets combined with the boundary state monitors. Aggregating across the nine sets gives us a 'validation' dataset, with validation predictions for each observation based on a fitted model that did not include the observations from that monitor. Finally, fitting the model to all the data except the tenth test set, we have a training dataset with nine-tenths of the data and a test set.

We consider predictive performance based on cross-validation to be the best approach for comparing and evaluating models, as accurate predictions are the goal for input into the health analysis. Furthermore, given the difficulties in defining good model selection criteria for nonparametric smoothing models, cross-validation has the benefit of clear interpretation and transparency for communication with nonstatisticians.

By including all observations for a given monitor in the same cross-validation set, our cross-validation assesses the ability of the model to predict at new locations. For the application, predictions are made at the residences of nurses. This introduces bias into our cross-validation results because of differences in the spatial distributions of the monitoring sites and residences.

Indeed, particularly for $PM_{10}$, some monitors are located to measure areas with high pollution rather than for population exposure, but Yanosky et al. (2008a) report that prediction accuracy is better for the population exposure monitors than for monitors sited at hot spots or those with unknown siting purpose, suggesting our predictions are reasonable for locations where people live. A validation substudy with measured gold-standard exposures at the residences of a subset of nurses would be invaluable for assessing our predictions, but such data are not available and are generally difficult to obtain. Also note that we do not account for the additional uncertainty in going from a monthly average based on subsampled days to the true monthly average from all days in the month, as most of the monitoring sites do not report every day. Finally, our predictions and uncertainties do not account for the difference between ambient concentrations and true personal exposure.

3.2. *Results.*

3.2.1. *Overview.* Figure 2 shows long-term average predictions on the original scale for both $PM_{10}$ and $PM_{2.5}$ for the northeast US, while Figure 3 shows predictions within a metropolitan area. These figures highlight the local spatial heterogeneity introduced by the covariates in the model, which is somewhat more pronounced for $PM_{10}$ than for $PM_{2.5}$. The models estimate 205 and 80 degrees of freedom (df) for $g_\mu(\cdot)$ for $PM_{10}$ and $PM_{2.5}$ respectively, and an average of 83 and 72 df over time for $g_t(\cdot)$ for $PM_{10}$ and $PM_{2.5}$, indicating that, particularly for the non-time-varying component, the model captures more spatial heterogeneity (after accounting for covariates) for $PM_{10}$. The residual variance for $PM_{10}$ is larger, as we would expect since $PM_{10}$ is more heterogeneous and by definition has higher concentrations, with time-invariant residual variance, $\hat{\sigma}_\mu^2$, of 0.021 for $PM_{10}$ and 0.0074 for $PM_{2.5}$, while the time-varying variances, $\hat{\sigma}_t^2$, are, on average over time, 0.023 for $PM_{10}$ and 0.013 for $PM_{2.5}$. The residual variance is generally larger in the winter than in the summer, with fall and spring intermediate, indicating more unmodeled heterogeneity in the winter. The regression terms are generally smooth, using fewer than 8 df, and most are close to linear.

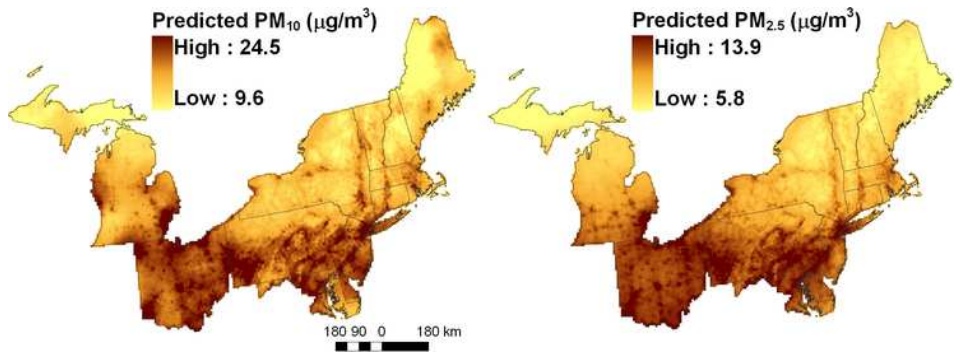

FIG. 2. *Maps of predicted $PM_{10}$ (left) and $PM_{2.5}$ (right) from the core model over the study region, averaged across all month-specific predictions from 1988–2002 for $PM_{10}$ and 1999–2002 for $PM_{2.5}$.*

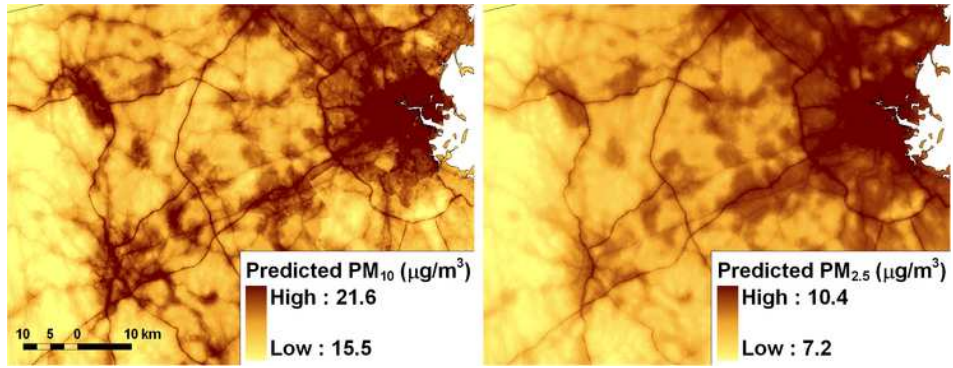

Fig. 3. *Map of predicted* $PM_{10}$ *(left) and* $PM_{2.5}$ *(right) in March 2000 for greater Boston, showing local heterogeneity. The dark area in the upper right of the plots is downtown Boston, while the two lines ringing the city reflect high predicted concentrations along the interstate beltways.*

### 3.2.2. *Prediction accuracy and coverage.*

*Monthly predictions.* Table 1 reports prediction accuracy on the training, validation, and test sets, based on the prediction $R^2$, $1 - \sum(y_{it} - \hat{y}_{it})^2 / \sum(y_{it} - \bar{y})^2$ and mean squared prediction error (MSPE), $\sum(y_{it} - \hat{y}_{it})^2/n$. Note that for $PM_{10}$, we exclude from the validation set an outlying site in Philadelphia with anomalously high concentrations and from the test set an anomalous site in northern Maine at the Canadian border for which the model predictions are based on spatial extrapolation from the training set and are much lower than the observations.

The models give reasonably high $R^2$, with the $PM_{2.5}$ model explaining more of the variability in the observations than the $PM_{10}$ model, in concordance with more spatial smoothness in $PM_{2.5}$. The use of the unbiased estimator generally has little effect on accuracy. The model overfits to some degree, which we suspect is associated primarily with the spatial surfaces, in particular, the monthly spatial surfaces, which we attempt to fit based on relatively sparse and noisy data. One avenue for further exploration would be to force more smoothness in the spatial terms to limit overfitting. Reassuringly, there is only a small difference between the validation results and the results on the test data, the gold standard as these data were held out of the entire modeling process. This suggests that the model selection process itself has not resulted in overfitting and that prediction results from the validation set, which is much larger than the test set, reflect generalizability to new locations. We also note that results for the monitors sited for population exposure are reasonably similar to those for all monitors, suggesting that our results can be generalized to the residential locations. The lower $R^2$ values seen in some cases for the population exposure monitors occur despite corresponding decreases in MSPE because of reduced PM variability at the population exposure monitors. The lack of everyday monitoring hinders our predictive modeling and assessment of the predictions; some of our inability to predict relates to the held-out observations being a subsample of days in the month.

TABLE 1
*Prediction accuracy and coverage of 95% prediction intervals for $PM_{10}$ and $PM_{2.5}$ for various subsets of the data. For $PM_{10}$ the validation and test sets exclude one site*

| | | Prediction accuracy: $R^2$(MSPE) | | |
|---|---|---|---|---|
| Data subset | log scale: $\hat{Y}_{it}$ | Original scale: $\exp(\hat{Y}_{it})$ | Original scale, unbiased: $\exp(\hat{Y}_{it} + \frac{1}{2}\widehat{\text{Var}}(\hat{Y}_{it}))$ | Coverage |
| $PM_{10}$ Training set | 0.815 (0.034) | 0.716 (32.5) | 0.721 (31.9) | 0.979 |
| Validation set | 0.668 (0.061) | 0.618 (40.7) | 0.616 (40.9) | 0.943 |
| Test set | 0.665 (0.059) | 0.615 (47.8) | 0.626 (46.4) | 0.940 |
| Validation set (pop'n exp.) | 0.646 (0.050) | 0.625 (31.7) | 0.607 (33.3) | 0.958 |
| $PM_{2.5}$ Training set | 0.872 (0.018) | 0.858 (3.0) | 0.860 (2.9) | 0.971 |
| Validation set | 0.770 (0.032) | 0.771 (5.0) | 0.771 (5.0) | 0.928 |
| Test set | 0.766 (0.033) | 0.742 (6.0) | 0.742 (6.0) | 0.914 |
| Validation set (pop'n exp.) | 0.731 (0.030) | 0.762 (4.5) | 0.760 (4.6) | 0.936 |

There was little obvious spatial pattern in MSPE by monitor, although for $PM_{2.5}$ some monitors in northern New England showed less accuracy (see explanation below). Not surprisingly, MSPE on the original PM scale was much higher for data points showing the highest observed concentrations, particularly for $PM_{10}$, as both the log transformation and the model smoothing cause predictions to be attenuated relative to the largest observations. However, MSPE was not noticeably lower for the lowest observed concentrations, and did not vary systematically with population density, suggesting predictive accuracy is similar between rural, suburban, and urban sites on an absolute basis, although the very rural IMPROVE sites did show lower MSPE than the AQS sites.

There was little evidence of systematic bias, except that predictions at the IMPROVE sites, which are in wilderness and park areas, had a bias of 1.73 and 0.52 for $PM_{10}$ and $PM_{2.5}$ respectively for IMPROVE, compared to $-0.58$ and $-0.20$ for AQS.

Coverage of 95% prediction intervals (calculated on the log scale and transformed as necessary) is good for $PM_{10}$ and slightly low for $PM_{2.5}$ (Table 1). This undercoverage presumably reflects several factors in such a complicated dataset and analysis, including non-normal error structure, heterogeneous data that are not fully characterized by the model, and estimates of standard errors that do not fully take dependence between the model components in (3.3) into account. Most sites had good coverage but a small number had very poor coverage (4 of 408 and 3 of 219 sites had coverage less than 50% for $PM_{10}$ and $PM_{2.5}$, resp.), with poor coverage occurring preferentially in northern New England and during cold months, suggesting the model has difficulty in capturing spatial variability in PM caused by wood stove smoke.

*Long-term predictions.* We consider the accuracy of long-term predictions, in particular, the average over 1988–2002 for $PM_{10}$ and 1999–2002 for $PM_{2.5}$, to assess the ability of the model to estimate cross-sectional heterogeneity (Table 2). Performance is even better than the monthly predictions, which is of particular interest for epidemiological studies assessing long-term health effects. The small number of test sites (35) may explain the poorer performance on the test set for $PM_{2.5}$. Prediction accuracy when exponentiating before averaging was very similar, albeit slightly lower. Coverage is reasonable, with overcoverage for $PM_{10}$, suggesting that the time-invariant uncertainty terms are able to capture the uncertainty common to all predictions over time at a given site. Finally, we considered the ability of the model to predict differences in long-term average PM between different locations, as the differences in exposure are what allow one to estimate health effects in epidemiological application. The model predicts differences in PM between monitoring sites very well. Differences of held-out observations plotted against differences of predictions follow the 1 :1 line, with a cross-validation $R^2$ for the differences of 0.72 for $PM_{10}$ and 0.81 for $PM_{2.5}$. Note that all the long-term average results are for monitors reporting at least 150 out of 180 possible monthly values for $PM_{10}$ and 39 out of 48 for $PM_{2.5}$, so a small amount of temporal variability induced by the missing months is reflected in the results. However, the results were largely insensitive to the cutoff used, indicating that they primarily reflect spatial heterogeneity.

TABLE 2
*Prediction accuracy and coverage of 95% prediction intervals for long-term averages of $PM_{10}$ (1988–2002; at least 150 months with observations) and $PM_{2.5}$ (1999–2002; at least 39 months) for various subsets of the data. For $PM_{10}$ the validation and test sets exclude one site*

|  | Data subset | Prediction accuracy: $R^2$(MSPE) | | Coverage |
|---|---|---|---|---|
|  |  | log scale: $\bar{\hat{Y}}_i$ | Original scale: $\exp(\bar{\hat{Y}}_i)$ |  |
| $PM_{10}$ | Training set | 0.846 (0.014) | 0.786 (10.16) | 0.982 |
|  | Validation set | 0.799 (0.021) | 0.722 (15.1) | 0.971 |
|  | Test set | 0.763 (0.022) | 0.718 (16.68) | 0.978 |
|  | Validation set (pop'n exp.) | 0.695 (0.018) | 0.681 (9.81) | 0.982 |
| $PM_{2.5}$ | Training set | 0.914 (0.0056) | 0.879 (1.09) | 0.977 |
|  | Validation set | 0.830 (0.011) | 0.804 (1.76) | 0.923 |
|  | Test set | 0.844 (0.015) | 0.764 (2.95) | 0.824 |
|  | Validation set (pop'n exp.) | 0.723 (0.012) | 0.763 (1.54) | 0.932 |

*Simplified and other alternative models.* Yanosky et al. (2008a, 2008b) compare the core model (3.1)–(3.2) with models that do not include covariates, models with more coarse time resolution in place of the monthly-varying surfaces, $g_t(\cdot)$, and simpler approaches to capturing the spatial structure, including inverse-distance-weighted interpolation and nearest neighbor methods. The core model substantially outperforms simpler models in terms of prediction accuracy, demonstrating the usefulness of including spatial, spatio-temporal, and regression structure in the model. Important covariates in the models include distance to the largest roads, elevation, wind speed, land use, and emissions.

Within the context of the models discussed here, we compared the core model with models with only linear terms in place of the smooth terms, $h_k(\cdot)$ and $f_j(\cdot)$, in (3.1)–(3.2) and without the covariates entirely, leaving only the spatial terms. Using linear terms in place of smooth regression terms led to minor decreases in predictive ability, with the cross-validation $R^2$ decreasing by less than 0.01 on the original and transformed scales for both $PM_{10}$ and $PM_{2.5}$, with the exception of long-term predictions of $PM_{10}$, for which the $R^2$ decreased from 0.799 to 0.780 and 0.722 to 0.694 on the log scale and original scales respectively. This suggests that with our rich set of covariates, smoothing may not be necessary, perhaps because the number of covariates provides a great deal of model flexibility simply from linear combinations of the covariates. In contrast, omitting the covariates entirely and relying purely on spatial smoothing greatly decreased predictive power, particularly for long-term averages. On the original scale, the cross-validation $R^2$ decreased from 0.618 to 0.558 for monthly $PM_{10}$, 0.722 to 0.559 for long-term $PM_{10}$, 0.771 to 0.717 for monthly $PM_{2.5}$, and 0.804 to 0.609 for long-term $PM_{2.5}$. This decrease in predictive ability suggests that the full model is using the covariates to capture substantial local heterogeneity that cannot be captured purely by spatial smoothing. Use of the covariates can also help to reduce the effective dimensionality of the model by attributing more variation to the low-dimensional covariate smooths and less to the spatial terms, with fewer resulting effective df used in the spatial smooths.

Our use of the penalized thin plate spline representation of the spatial surfaces was one of several possibilities to represent the spatial structure in the data. This choice allows us to fit the model with gam and to fit the second stage model in one step. In the supplementary material [Paciorek et al. (2009), Section S1], we consider alternative statistical specifications for the spatial and regression terms in the model, including kriging, reporting that none of the alternatives improved upon the predictive performance of our core model.

*Coarse PM.* To estimate coarse PM, we took the simple approach of differencing the predictions for $PM_{10}$ and $PM_{2.5}$. We carried out a small cross-validation exercise for 1999–2002 in which we sequentially held out one-tenth of the monitors that report both $PM_{2.5}$ and $PM_{10}$. The cross-validation $R^2$ of 0.370 (MSPE of 32.3) indicates that this approach is less successful in predicting monthly coarse PM than our models for $PM_{10}$ and $PM_{2.5}$. This is not surprising given the physical reasons for greater heterogeneity in coarse PM than fine PM and the fact that the differenced observations contain two instrument error components. Training error ($R^2$ of 0.622 and MSPE of 19.36) is much less than cross-validation error, suggesting substantial overfitting, while cross-validation error is no better for population exposure monitors. In contrast, the approach does a good job of predicting long-term (1999–2002 for sites with at least 39 months of data) average coarse PM, with cross-validation $R^2$ of 0.619 (MSPE of 10.32) after excluding the same northern Maine site excluded from the $PM_{10}$ validation.

An alternative to our approach would be to fit a model to differences in the observed $PM_{10}$ and $PM_{2.5}$ values, but co-located monitors represent only about half of the monitors during 1999–2002 [Paciorek et al. (2009), Figure S1]. In contrast, a joint modeling approach would be able to use all the data and provide a principled approach to estimate uncertainty, potentially at the daily scale with the goal of better monthly predictions. However, this would involve the added complexity and potential biases induced by modeling the relationship between $PM_{10}$ and $PM_{2.5}$, so we have not pursued that approach in this project.

### 3.2.3. *Residual structure.*

*Normality.* Our model (3.1)–(3.2) assumes normal, homoscedastic errors for both the time-varying first-stage residuals and second-stage site-specific effects. For $PM_{10}$, the time-varying residuals from the first stage, standardized by the month-specific residual variances, are somewhat right-skewed and have a long tail. For $PM_{2.5}$, the residuals are reasonably symmetric, but with long tails, approximated by a $t$ distribution with 5 df. While some sites have residuals with long tails, many sites have residuals that are well fit by a normal distribution. The residuals in the second stage model, $\hat{\mu}_i - \hat{g}_\mu(s(i)) - \sum_j \hat{f}_j(z_{ji})$, also have somewhat long tails.

Smith et al. (2003) found a square root transformation rather than a log transformation to be better for $PM_{2.5}$ data in the southeastern US. We fit a model with the square root transformation for both $PM_{10}$ and $PM_{2.5}$ and found that the residual structure was similar to that under the log transformation. The validation predictive accuracy on the original scale was slightly better using the square-root transformation, 0.623 and 0.784 for $PM_{10}$ and $PM_{2.5}$ respectively, compared to 0.618 and 0.771 for the log transformation.

*Residual autocorrelation.* Figure 4(a), (b) shows the residual autocorrelation from the first-stage model. There is some autocorrelation at short lags and some seasonal structure, which is not surprising given that we have not smoothed in time, except to the extent accounted for by time-varying covariates, primarily wind speed. The relatively small magnitude of the autocorrelation is perhaps not surprising since we have aggregated the data to the monthly level.

To assess whether the systematic temporal pattern in the residuals represents information that could help to improve our predictions, we examined semivariograms of the model residuals, which showed no evidence of spatial pattern (Figure 5). Individual locations have correlated residuals over time, but these residuals are not correlated with the residuals at nearby monitors. We believe this reflects local spatial heterogeneity that is not useful for predicting PM except in the very local vicinity of the monitors. Gneit-ing (2002) refers to this phenomenon of temporal autocorrelation without spatial autocorrelation as a spatial nugget.

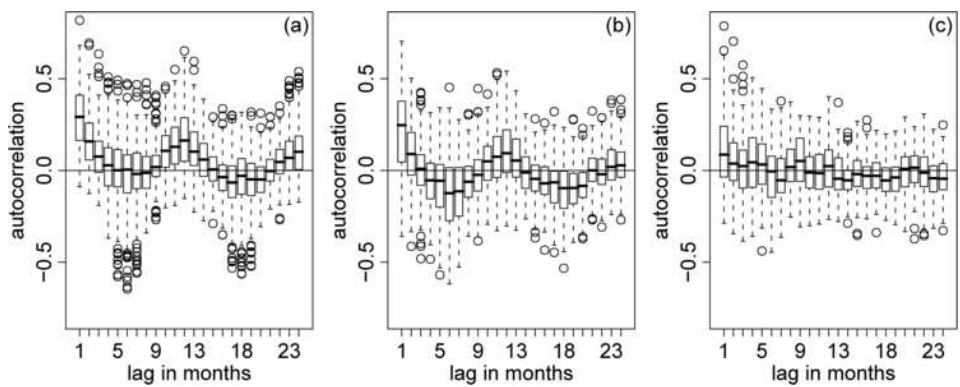

FIG. 4. *Residual temporal autocorrelation for models of the form (3.1)–(3.2) for (a) $PM_{10}$ for 1988–2002 and (b) $PM_{2.5}$ for 1999–2002. Panel (c) shows the residual temporal autocorrelation of the log ratio model for $PM_{2.5}$ for 1988–1998 (4.1). Each point in a given boxplot is the average lag-correlation at one site, using only sites with at least 100 observations in (a) and 30 observations in (b) and (c).*

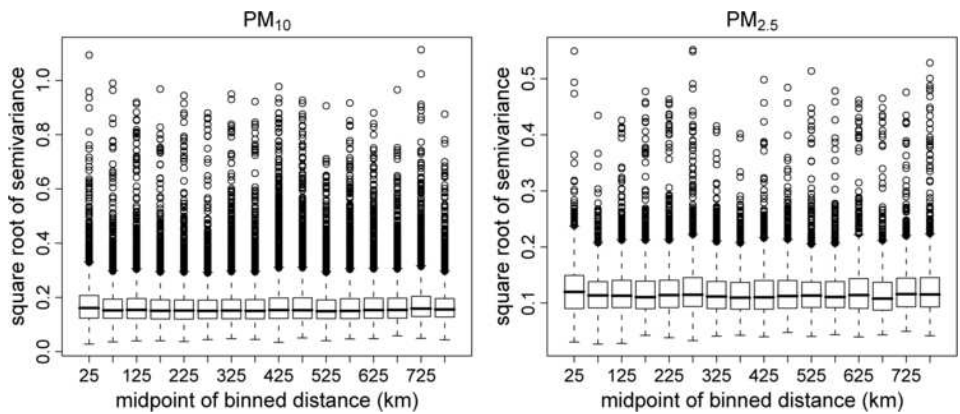

FIG. 5. *Residual spatial semivariance plots for $PM_{10}$ (left) and $PM_{2.5}$ (right) models of the form (3.1)–(3.2), with each point in a given boxplot the square root of the average squared difference (of observations co-occurring in time) between two sites. Pairs of sites are binned based on the distance between them, and we use only pairs with at least 10 co-occurring observations.*

Consistent with this, when we included short-term lag structure [Wood (2006), pages 162–167] and seasonal spatial terms to extend (3.1)–(3.2), we found some reduction in residual autocorrelation, but little or no improvement in predictive performance, with a large cost in computational speed. Models using a full space–time covariance structure would be computationally difficult, particularly since the set of monitors with data changes over time, obviating the computational advantages of a separable covariance.

3.3. *Epidemiological analysis.* Puett et al. (2008) report epidemiological results for $PM_{10}$ using predictions from the core model (3.1)–(3.2) in a prospective cohort of 66,250 women from the Nurses' Health Study in northeastern US metropolitan areas. Nurses' addresses and personal information were obtained by questionnaire starting in 1976 and updated every two years. Addresses were geocoded to obtain latitude and longitude. Cox models were run at the monthly level with stratification by age in months. The covariates were year, season, smoking status, pack years of smoking, family history of myocardial infarction, body mass index, presence of high cholesterol, presence of diabetes, presence of hypertension, median family income in census tract of residence, level of physical activity, and median house value in census tract of residence (a proxy for wealth). State of residence was also included as a fixed effect to help control for confounding by unmeasured variables that may vary spatially, helping to account for regional-scale variation in health outcomes. Puett et al. (2008) focused on the association between health outcomes and $PM_{10}$ exposure when exposure was estimated as the average over the previous 12 months, so we consider that averaging period here. The health outcomes most strongly associated with $PM_{10}$ were all-cause mortality (excluding accidents) and incident fatal coronary heart disease (CHD). Results varied somewhat depending on the covariates included in the analysis.

TABLE 3
*Hazard ratios (95% confidence intervals) with a 10 $\mu g/m^3$ change in 12-month average $PM_{10}$ for all-cause mortality and fatal CHD using three different sets of exposure predictions. All estimates are based on Cox models at the monthly level, stratifying by age in months and including a number of covariates. Results from the core model were reported in Puett et al. ([2008](#))*

| Exposure predictions used | All-cause mortality | Fatal CHD |
|---|---|---|
| Core model [(3.1)](#)–[(3.2)](#) | 1.07 (0.97, 1.18) | 1.30 (1.03, 1.77) |
| Space–time model without covariates | 1.07 (0.97, 1.18) | 0.92 (0.69, 1.21) |
| Nearest monitor | 0.98 (0.93, 1.05) | 0.90 (0.76, 1.07) |

As a sensitivity analysis, we used estimates of exposure from simpler approaches in place of predictions from the core model. In particular, we compared the previous results with use of the average 12-month (Jan–Dec) PM concentration for the year 2000 from the nearest monitor reporting at least 9 of the 12 months, excluding nurses further than 50 km from the nearest monitor, an approach similar to that of Miller et al. ([2007](#)). To estimate exposure at a given time based on the most recent known address, we used the concentration in the year 2000 from the nearest monitor to that address. We also used the previous 12-month average from our two-stage space–time model without the GIS and meteorological covariates, which has the flavor of a simple kriging model, but with the overall spatial term included as the second stage model (Section [3.2.2](#)). Table [3](#) shows the results for mortality and fatal CHD. With the exception of all-cause mortality for the simple space–time model without covariates, results were quite different from the core analysis, with attenuated effect estimates consistent with the null hypothesis. This suggests the importance of improved exposure estimates. The equivalence of results for all-cause mortality between the core model and the simple space–time model was surprising given the fine-scale heterogeneity in exposure modeled in the full model and the reduction in predictive ability of the simple model compared to our full model (Section [3.2.2](#)). Similar analyses for $PM_{2.5}$ are in submission in the applied literature. An important advantage of our modeling approach is the ability to estimate exposure for any time lag of interest in a simple way that accounts for residential movement and the actual dates of the health outcomes.

The prediction error in our model becomes measurement error with implications for bias when using the predictions in a health analysis. In the supplemental material [Paciorek et al. (2009), Section S2] we argue that the exposure modeling takes the form of regression calibration with the implication of limited bias in health analyses [Gryparis et al. (2009)]. However, the assessment does leave aside sources of error we cannot quantify that may reflect classical measurement error, in which the variable as measured is a noisy version of its true value [see Paciorek et al. (2009), Section S2].

**4. Modeling sparse $PM_{2.5}$ data conditional on $PM_{10}$ and visibility.** For the period 1988–1998, there were few $PM_{2.5}$ monitors reporting data, with many of the monitors located in rural and protected areas (from the IMPROVE network). Hence, our model for this period relies on proxies for $PM_{2.5}$, in particular, the predicted $PM_{10}$ from our core $PM_{10}$ model and airport visibility information. Our basic approach is to model $PM_{2.5}$ conditionally on $PM_{10}$, using $PM_{10}$ as a covariate, rather than in a joint model. Conditional modeling is simpler and allows us to use gam and follow the same basic approach used in Section 3.1.1. Also, because $PM_{10}$ monitors are much more abundant than $PM_{2.5}$ monitors during this period, our goal is to leverage the $PM_{10}$ data to improve $PM_{2.5}$ predictions, with little gain seen from a joint model.

4.1. *Model.* Models for $PM_{2.5}$ for 1988–1998 were fit using available $PM_{2.5}$ data from 1988–2002 as well as predictions from the $PM_{10}$ model for all relevant locations and months and airport visibility information. The visibility measure, $b_{\text{ext}}$, is constructed and predicted at all locations for all months as described in Section 4.2.

The final model chosen was a model that fits the ratio of $PM_{2.5}$ to predicted $PM_{10}$, using covariates and borrowing strength across space to estimate the fraction of $PM_{10}$ that is fine particulate matter, $PM_{2.5}$. $PM_{2.5}$ is a subcomponent of $PM_{10}$, so knowledge of $PM_{10}$ provides a great deal of information about $PM_{2.5}$. The model takes a similar form to the core two-stage model, but in the first stage, we model the log of the ratio,

$$
\begin{aligned}
y_{it} &= \log \frac{PM_{2.5,it}}{\widehat{PM}_{10,it}} \\
(4.1) \quad &\sim \mathcal{N}\bigg(\mu_i + m(t) + g_{\text{season(t)}}(s(i)) + h_{PM_{10}}(\log \widehat{PM}_{10,it}) \\
&\qquad\qquad + h_{\text{vis}}(\log \hat{b}_{\text{ext},it}) + \sum_k h_k(x_{kit}), \sigma^2_{\text{season(t)}}\bigg),
\end{aligned}
$$

while the second stage is identical to the second stage of the core model (3.2), but with different covariates. An additional covariate is the visibility information, $\hat{b}_{\text{ext}}$ (see Section 4.2). Note that we also include predicted $PM_{10}$ as a covariate, as the PM ratio may vary based on the overall level of $PM_{10}$. In using the logarithm of the ratio, we can move $\log \widehat{PM}_{10,it}$ to the right-hand side, so we see that we have a model for $\log PM_{2.5}$ that includes a fixed offset for $\log \widehat{PM}_{10}$. We fit separate spatial surfaces for the four seasons rather than for individual months because the data cannot inform monthly surfaces for the whole time period. By fitting the model to $PM_{2.5}$ data from the entire period, 1988–2002, we rely primarily on the more numerous 1999–2002 data to model the basic spatial patterns and covariate effects in the PM ratio, which are then assumed to hold before 1998. We believe that this assumption of constancy over time is a reasonable assumption for the PM ratio, particularly for the northeast US. Estimation of the time trend component, $m(t)$, modeled as a penalized spline term, which relies on data throughout the period, allows the model to capture a purely temporal trend in the ratio, although there is little evidence of overall trend from the small number of co-located monitors. Yanosky et al. (2008b) describe several alternative models that use different time resolutions for the spatial surfaces and different definitions of the outcome variable; the model described here outperformed the alternative approaches.

We can make predictions at arbitrary times and locations using the predicted log ratio as $\widehat{PM}_{2.5,it} = \exp(\hat{y}_{it})\widehat{PM}_{10,it}$, which tells us how to scale the $PM_{10}$ prediction at each location to predict $PM_{2.5}$. In some cases we predict $PM_{2.5}$ to be larger than predicted $PM_{10}$. While this is physically impossible, we believe the model structure is able to give reasonable predictions, despite situations in which the $PM_{10}$ and $PM_{2.5}$ predictions are inconsistent.

$PM_{2.5}$ monitoring data before 1999 are sparse in space and time, with monitors available at a small set of locations, biased toward rural areas because of the IMPROVE network (23% of observations, compared to 5% in 1999–2002) and toward high concentration locations for the available AQS monitors (36% of observations compared to 11% in 1999–2002) and with each monitor reporting for haphazard periods of time. Cross-validation is more difficult because the unusual characteristics of the $PM_{2.5}$ monitors available before 1999 may introduce bias into the cross-validation assessment. We take two cross-validation approaches to assess the model. First we fit the model using 1988–2002 data, holding out and making predictions for most of the data from 1999. This allows us to see how well the model does in predicting $PM_{2.5}$ under cross-validation with a large amount of held-out data. However, because 1999 is only one year lagged from 2000, this is an easier task than making predictions for 1988–1998, for which seasonal and spatial patterns may have changed and with more of an influence of any ongoing time trend. Therefore, we also divided the 1988–1998 data into five sets and held out each in turn for cross-validation. However, the limited amount of data makes it difficult to ensure that results are robust, and the locations of monitors in that period are not spatially representative (Figure 1b).

4.2. *Modeling visibility.* Visibility is known to be related to fine PM concentrations under clear weather conditions [Ozkaynak et al. (1985)], with PM haze reducing visibility. Measurements of visible range collected over time at airport weather stations have been used as a surrogate measure to estimate $PM_{2.5}$ at various locations in the United States. However, observations of visual range have several shortcomings. The visible range measurement is the distance to the most distant marker that is visible among a set of permanent markers at fixed intervals. Often this marker is the furthest marker, so the visibility data are interval-censored and right-truncated. In the mid-1990s, visibility measurements were automated, resulting in a maximum visible range of 10 km, which corresponds to extremely high PM concentrations, greatly reducing the informativeness of the data as a proxy for PM. The relationship between visibility and particulates varies by humidity in the atmosphere, and any precipitation or fog prevents use of the visibility data. Past work has involved ad hoc approaches to manipulate the visibility data for use in estimating $PM_{2.5}$ [Ozkaynak et al. (1985)].

We use smoothing and truncation techniques in a statistical model to construct a visibility proxy for $PM_{2.5}$ by estimating relative humidity (RH)-adjusted beta extinction, $b_{\text{ext}}$, a measure of light extinction, as a proxy for $PM_{2.5}$, at all required spatial locations for use in (4.1). Estimates of beta extinction can be calculated from visual range data using the Koshmeider formula, $b_{\text{ext}} = \frac{K}{V}$, where $K$ is the Koshmeider constant (unitless), $V$ is the visual range in kilometers, and $b_{\text{ext}}$ is the estimated beta extinction in inverse kilometers. Investigators have compared estimates of beta extinction derived from visual range in the US and have found a value, $K \approx 1.9$, that results in estimates of beta extinction that are correctly scaled to observations of $PM_{2.5}$ [e.g., Ozkaynak et al. (1985)]. Consistent with previous studies, we discard visibility observations made during periods of precipitation as recorded at the airport weather stations, as well as those where visibility is very low (which is indicative of fog or precipitation, rather than extreme pollution) or where the RH is very high (above 99%; indicative of fog).

4.2.1. *Calibration to 60% relative humidity.* To adjust for the effect of RH on the beta extinction estimates, we used all the observations over space and time to regress against a penalized spline smooth term of RH using `gam` for each season (winter, spring, summer, fall). In the regression we used log-transformed bounds on beta extinction where the bounds were calculated based on the interval censoring-induced bounds on visual range. The log transformation was used because of the approximately lognormal distribution of daily beta extinction values. To remove the effect of Raleigh scattering on the beta extinction estimates, a constant value of 0.10 $\text{km}^{-1}$ was subtracted from the bounds. In the fitting, we account for the censoring of the beta extinction values using a stochastic expectation-maximization (EM) algorithm because the penalized spline regression model does not provide us with a simple closed-form expected likelihood to maximize. We start by regressing the midpoint of the lower and upper bounds of $\log b_{\text{ext}}$ on RH using gam. Then, for each observation, taking the expected value of $\log b_{\text{ext},it}$ from the model fit, we sample $\log b_{\text{ext},it}^{(k)}$ from a truncated normal distribution and proceed by iterating between fitting using gam with the sample values and sampling the values. We iterate until convergence (200 total iterations, with convergence after 20–30 iterations), taking the mean of the last 100 iterations on a fine grid of RH values as our calibration curve for a given season. We correct the lower and upper bound beta extinction values at each station to 60% RH using an additive adjustment based on the fitted seasonal calibration models. Note that we assume that the fitted calibration curve is known, but given the large amount of data, this assumption seems reasonable. One area of potential concern is that if pollution levels vary systematically with average RH, we may be calibrating some of the influence of pollution on beta extinction out of the relationship.

4.2.2. *Spatial smoothing of beta extinction.* The above procedure gives us RH-corrected lower and upper bounds for beta extinction for each day (at mid-day) at each airport station reporting visibility. We spatially smooth the airport RH-corrected beta extinction for each day and make predictions at the locations of interest. We then use the average across the days in each month at each location for use in (4.1). The spatial smoothing needs to account for the interval-censored nature of the RH-corrected beta extinction. We again use a stochastic EM approach in which we iterate between smoothing sampled values and sampling values from a truncated normal distribution with mean from the current fitted model and truncation limits based on the bounds of the RH-corrected beta extinction values. In this way, beta extinction is predicted at any location within the spatial domain of the data using information from nearby locations on a daily basis.

4.3. *Results.*

4.3.1. *Overview.* Long-term average predictions for the 1988–1998 period (not shown) are similar to those in Figure 2b. The model estimates more spatial variability in the log ratio in the summer and winter, with 303 and 355 estimated degrees of freedom for the $g_{\text{season}}(s)$ terms in the summer and winter, compared to 38 and 5 degrees of freedom for spring and fall, respectively. These surfaces indicate substantial spatial variability in the ratio of $PM_{2.5}$ to $PM_{10}$, as does the time-invariant spatial surface, with 70 estimated degrees of freedom. This spatial variability and the substantial covariate effects in the log ratio model support the use of the model rather than just using $PM_{10}$ as a simple surrogate for $PM_{2.5}$, as do results in Yanosky et al. (2008b). There is no apparent trend over time in the log ratio as indicated by the fitted $m(t)$, but there are clear seasonal patterns, with peaks during winter. Variograms of the model residuals indicate no evidence of spatial pattern, providing some support for our assumption of constancy in the spatial pattern of the ratio over time.

4.3.2. *Prediction accuracy and coverage.* Table 4 shows prediction accuracy on the $PM_{2.5}$ scale, comparing $\widehat{PM}_{2.5}$ with $PM_{2.5}$ observations. Performance is generally good, although there is overfitting, as in the core model. We have excluded one site from the pre-1999 validation set, a site reporting high concentrations in the New York City area for which the model substantially underpredicts. Note that the pre-1999 data have more variability, which helps to enhance predictability, as seen in the higher $R^2$ accompanied by higher MSPE.

One concern with the validation results is that these include comparisons at locations with co-located $PM_{10}$ monitors. We would expect the model to do better at these locations than elsewhere because the predictions from the $PM_{10}$ model are likely to be better where there is a monitor, and these $PM_{10}$ predictions are an important part of the log ratio model. Excluding sites with a co-located $PM_{10}$ monitor, in both time periods our prediction assessments do not appear to be overly optimistic, with an $R^2$ of 0.636 (MSPE of 11.0) for the 1988–1998 set and 0.652 (MSPE of 7.4) for the 1999 set.

On the scale of $PM_{2.5}$, bias is positive, but reasonably small, for both the 1988–1998 validation data, excluding the site mentioned above, at 0.49 relative to the standard deviation of the observations of 8.61 and on the 1999 validation data, with bias of 0.37 and a standard deviation of 4.90.

Table 4 indicates that coverage is good for the training set and 1999 validation set, despite the simplifications used in estimating standard errors (Section 3.1.1). Coverage on the validation set for 1988–1998 is poor, indicating the model has some difficulty, strongly underestimating uncertainty, presumably in part because of unusual characteristics of the pre-1999 monitors.

TABLE 4
*Accuracy and coverage for various subsets of the data from the log ratio model. The 1988–1998 validation set excludes one site*

|  | $R^2$(MSPE) | Coverage |
|---|---|---|
| Training set | 0.753 (8.6) | 0.953 |
| Validation set (1988–1998) | 0.659 (25.2) | 0.831 |
| Validation set (1999) | 0.641 (8.6) | 0.924 |
| Validation set (1999; pop'n exp.) | 0.630 (7.4) | 0.928 |

The residuals were symmetric but long-tailed, as in Section 3.2.3, while the temporal autocorrelation was minimal (Figure 4c).

**5. Discussion.** We have described an extensive modeling effort for PM used to predict concentrations as input for health analysis in a large cohort study. From a statistical perspective, our model represents a trade-off, incorporating covariate effects and spatio-temporal structure in a two-stage model but with a somewhat simplified structure. We believe the model appropriately balances accounting for key factors affecting heterogeneity in concentrations with retaining computational feasibility and transparency for scientific communication. The modeling is being extended to other parts of the US to add exposure estimates for a larger portion of the NHS cohort, and we hope to use it for other cohorts as well.

The model borrows from a variety of spatio-temporal and semiparametric regression approaches in the literature and captures key aspects of the data that are not captured in other approaches. The two-stage approach is similar in spirit to the ANOVA space–time decomposition of Diez-Roux et al. (in preparation). Their first-stage ANOVA decomposes the variability into site- and time-specific fixed effects. The site-specific effects are then modeled as random variables conditional on covariates, but they do not consider spatio-temporal interaction, as represented in our monthly spatial surfaces. The form of the model with a time-invariant spatial surface and space–time residuals modeled independently at each time increment is similar to Smith et al. (2003), but we deal with a much larger set of covariates to represent fine-scale spatial heterogeneity and include the site-specific intercept and second-stage model to help avoid overfitting.

Yanosky et al. (2008a, 2008b) demonstrate that the model outperforms simpler specifications. The alternative statistical formulations we present in the supplementary material [Paciorek et al. (2009), Section S1] did not improve predictive performance markedly. With respect to the key assumption of independence of the spatial residual surfaces over time, residual diagnostics indicate some residual autocorrelation in time, but no autocorrelation in space, suggesting there is no additional spatial information in the residuals that would improve our predictions and supporting our analytic arguments (Section 3.1.1). Other specifications might better account for the autocorrelation but this appears to not be warranted from the perspective of model predictive performance, is difficult computationally with such a large dataset, and introduces logistical difficulties to quick implementation of the model by environmental scientists for use in the health modeling.

Placement of monitors by EPA and the states was not based on statistical design considerations. The goals included standards attainment, convenience, and scientific understanding of PM processes, in addition to the goal of representing population exposure. Monitors are much more dense in urban areas than in suburban and rural areas. This creates the potential for biased estimates of exposure, most likely an upward bias because of dense monitoring in urban areas and monitors situated to capture high levels of PM. We assume that pollution levels at locations without monitors are missing at random (MAR) and therefore ignorable, conditional on the covariates we include in the model to represent local conditions. Heuristically, this assumes that the covariates allow the model to localize anomalous values such that their influence does not extend very far in space. For example, an anomalously high value may be accounted for by covariates reflecting high local emissions or by being near a major road, while not pulling up the residual spatial surface, so predictions nearby at locations with different covariate values may be much lower than at the monitor. The MAR assumption is undoubtedly not completely true, but we believe it represents an improvement over purely spatial models without site-specific covariates. In our model, inclusion of monitors sited to detect high concentrations and point source influences may positively bias predictions, but exclusion of such monitors may negatively bias predictions in the vicinity of point sources and hot spots and reduces spatial coverage.

While additional monitoring data are unlikely to become available in the future and are not available for retrospective estimation, other sources of proxy information, such as satellite remote sensing data [e.g., Liu et al. (2005)] and deterministic atmospheric chemistry models, may provide some information, especially in rural and suburban areas far from monitors and on days with few monitors reporting, but evidence of improved predictions is needed. Much current statistical interest lies in combining such sources of information [Fuentes and Raftery (2005), van de Kassteele and Stein (2006), McMillan et al. (2008)]. We believe that our work serves as a reasonable baseline model of PM concentrations that can be a rigorous point of comparison to judge whether models incorporating these additional sources of information improve predictions of PM. The challenge for the applied use of models with these additional data sources is specification of space–time models and fitting techniques that are computationally feasible for many time periods and large spatial domains. Work in this area is underway by the first author.

**Acknowledgments.** We thank Joel Schwartz, Jaime Hart, and Frank Speizer for discussions of the modeling as part of the larger project, Louise Ryan for feedback, Steve Melly for GIS processing, and the editor, Michael Stein, for helpful comments. Although the research described in this article has been funded wholly or in part by the United States Environmental Protection Agency through grant EPA STAR R83-0545-010 to Francine Laden, it has not been subjected to the Agency's required peer and policy review and therefore does not necessarily reflect the views of the Agency and no official endorsement should be inferred.

## SUPPLEMENTARY MATERIAL

Additional discussion of the measurement error implications of using model predictions in the health analyses and of modeling alternatives to the core model are available as supplementary material linked to the paper at the journal web site, as are the data used here and R code for fitting the core model.

**Supplement A: Supplementary discussion of alternative models and measurement error implications** (DOI: 10.1214/08-AOAS204SUPPA; .pdf). We first consider several alternative statistical specifications for the spatial and regression terms in the model, including kriging, concluding that none of the alternatives improve upon the predictive performance of our core model. Next we consider the measurement error implications of using the model predictions in an epidemiological analysis as a covariate, arguing that the exposure modeling takes the form of regression calibration with the implication of limited bias in health analyses [Gryparis et al. (2009)]. However, the assessment does leave aside sources of error we cannot quantify that may reflect classical measurement error.

**Supplement B: Supplementary code and data** (DOI: 10.1214/08-AOAS204SUPPB; .zip). R code for fitting the core model and the data used here.

C. J. Paciorek  
Department of Biostatistics  
Harvard School of Public Health  
Boston, Massachusetts 02115  
USA  
E-mail: paciorek@alumni.cmu.edu  
URL: http://www.hsph.harvard.edu/~paciorek

J. D. Yanosky  
H. H. Suh  
Department of Environmental Health  
Harvard School of Public Health  
Boston, Massachusetts 02115  
USA  
E-mail: jyanosky@hsph.harvard.edu  
hsuh@hsph.harvard.edu

R. C. Puett  
Cancer Prevention and Control Program  
Department of Epidemiology and  
Biostatistics and Department of  
Environmental Health Sciences  
Arnold School of Public Health  
University of South Carolina  
Columbia, South Carolina 29208  
USA  
E-mail: rpuett@gwm.sc.edu

F. Laden  
Channing Laboratory  
Brigham and Women's Hospital  
and Harvard Medical School  
Boston, Massachusetts 02215  
and  
Departments of Environmental Health  
and Epidemiology  
Harvard School of Public Health  
Boston, Massachusetts 02115  
USA  
E-mail: francine.laden@channing.harvard.edu